# The Inner Limit of Quantum Brownian Evolution and Its Relevance to Positivity

Allan Tameshtit

The conventional quantum Brownian propagator, which describes the evolution of a system of interest bilinearly coupled to and initially uncorrelated with a reservoir, does not preserve positivity of density operators, prompting workers to modify the propagator by the *ad hoc* addition of time-independent terms to the corresponding generator. We show that no such terms need be added to the generator to preserve positivity provided one accounts for the rapid entanglement of the system of interest and the reservoir on a time scale too short for the conventional propagator to be valid.

**Introduction**
The conventional quantum Brownian propagator for a harmonic oscillator, $\exp(tL)$, has the generator [1,2]

$$L = \frac{1}{\hbar i}[H,\cdot] + \frac{\Gamma}{2\hbar i}\{p,\cdot,q\} - \frac{\Gamma}{2\hbar i}\{q,\cdot,p\} - \frac{2\Gamma mkT}{\hbar^2}\{q,\cdot,q\}, \tag{1}$$

where $\Gamma$ is a positive coupling constant, $\{A,\rho,B\} \equiv BA^\dagger \rho + \rho BA^\dagger - 2A^\dagger \rho B$ and

$$H = \frac{p^2}{2m} + \frac{\Gamma}{2}(qp+pq) + \frac{1}{2}m\omega^2 q^2. \tag{2}$$

This propagator may be derived by considering a total system consisting of the harmonic oscillator coupled to a reservoir of other oscillators, where initially the harmonic oscillator and the reservoir are uncorrelated [1,2]. Although possessing many desirable features, such as a close correspondence to the classical analogue, it is known [3-5] that a problem can arise when using this propagator: $\exp(tL)$ does not map the set of density operators V (i.e., the set of positive Hermitian operators with unit trace) into itself [6]. In particular, if $\rho \in$ V, then $\exp(tL)\rho$ is of unit trace and Hermitian, but is not necessarily positive for all $t > 0$. Because a basic tenet of quantum mechanics is that density operators representing a system of interest be positive, $L$ has been modified in the literature by the *ad hoc* addition of the term $-\Gamma\{p,\cdot,p\}/8mkT$ to preserve positivity [7]; in fact, the addition to $L$ of any time independent term that is proportional to $\{p,\cdot,p\}$, where the constant of proportionality is less than or equal to $-\Gamma/8mkT$, ensures positive evolution.

In this paper, we show that no such *ad hoc* term need be added to $L$ to preserve positivity provided we account for the rapid entanglement of the harmonic oscillator and reservoir on a time scale $\Delta t$ too short for $\exp(tL)$ to be valid [8]. This fast evolution, which is associated with a boundary layer and arises in part because the harmonic oscillator and the reservoir to which it is coupled are initially uncorrelated, is described by an inner propagator $J(\Delta t)$ that acts on this short time scale $\Delta t$. We derive the inner propagator below from first principles in an inner limit. The inner propagator maps V to a smaller set V' of generally non-pure, positive density operators: V' =J(Δt) V ⊂ V.

The physical picture that emerges is that of a density operator entangling rapidly with the reservoir in an inner limit, and evolving subsequently with the approximate outer propagator $\exp(tL)$ in an outer limit, all the while remaining positive. (In the context of a two level system coupled to a reservoir, see also [9], which reference is discussed below.) A simple patch of the inner and outer limits yields positive evolution [10]:

$$\rho(0) \xrightarrow{J(\Delta t)} \rho(\Delta t) \xrightarrow{\exp((t-\Delta t)L)} \rho(t) \in V, \text{ for } t \geq \Delta t.$$

**Exact Solution**

Perhaps the most ubiquitous model leading to Equation (1) for studying quantum Brownian motion is that of a harmonic oscillator (system of interest) bilinearly coupled to an infinite number of other oscillators (reservoir). The total Hamiltonian describing such a system of interest and reservoir is

$$H_T = \sum_{v=0}^{N} \left( \frac{p_v^2}{2m_v} + \frac{m_v \omega_v^2 q_v^2}{2} \right) + \sum_{n=1}^{N} \varepsilon_n q_0 q_n, \qquad (3)$$

where $(q_0, p_0)$ and $(q_1,...,q_N, p_1,...,p_N)$ are the canonical coordinates of the system of interest and reservoir, respectively. The Liouville equation associated with $H_T$ was solved exactly in Ref. [2] in the representation that yields the total Wigner function, $W_T$. After introducing a spectral strength function

$$f(\omega) = \frac{2}{\pi} \frac{\kappa \alpha^2 \omega^2}{\alpha^2 + \omega^2}, \qquad (4)$$

where $\alpha$ is a high frequency cut-off, $\kappa$ is a measure of the coupling strength between the system of interest and the reservoir, and $\omega$ are the frequencies of the reservoir oscillators, and after assuming a factorized initial state, Reibold and Haake [2] when on to compute the exact, reduced Wigner function, $W(q, p, t)$, (obtained by integrating $W_T$ over the reservoir variables) describing the time evolution of the system of interest:

$$W(q,p,t) = \frac{1}{(2\pi)^2 \hbar} \int_{-\infty}^{\infty} dk \int_{-\infty}^{\infty} dk' \exp i \left( \sqrt{\frac{m\omega}{\hbar}} kq + k'p/\sqrt{\hbar m\omega} \right) \times$$

$$\tilde{W}_0 \left( k \dot{A}\sqrt{\omega/\hbar} + k' \ddot{A}/\sqrt{\hbar\omega}, kA\sqrt{\omega/\hbar} + k' \dot{A}/\sqrt{\hbar\omega} \right) \exp\left[ -\left( \omega X k^2/\hbar + Y k'^2/\hbar\omega + \dot{X} kk'/\hbar \right)/2 \right],$$

(5)

where $\tilde{W}_0(k,k')$ is the Fourier transform of the initial Wigner function,

$$A = \frac{2\Gamma[\exp((2\Gamma-\alpha)t) - \exp(-\Gamma t)\cos(\Omega t)] + \Omega^{-1}[(\alpha-2\Gamma)^2 + \Omega^2 - \Gamma^2]\exp(-\Gamma t)\sin(\Omega t)}{(\alpha-3\Gamma)^2 + \Omega^2},$$

with $\Omega^2 = \dfrac{\alpha\omega^2}{\alpha - 2\Gamma} - \Gamma^2$ and $\Gamma \sim \kappa/2$ as $\alpha \to \infty$,

$$X(t) = \frac{\hbar}{2}\int_0^\infty d\omega \frac{f(\omega)}{\omega} \left|\int_0^t dt' e^{i\omega t'} A(t')\right|^2 \coth\left(\frac{\hbar\omega}{2kT}\right), \tag{6}$$

and

$$Y(t) = \frac{\hbar}{2}\int_0^\infty d\omega \frac{f(\omega)}{\omega} \left|\int_0^t dt' e^{i\omega t'} \dot{A}(t')\right|^2 \coth\left(\frac{\hbar\omega}{2kT}\right). \tag{7}$$

However useful this solution is, the structure of $W(q, p, t)$ does not lend itself to easily discerning the underlying physics of the system of interest in contact with its surroundings. Moreover, the Wigner function is not well suited to theoretical studies related to positivity of the reduced density operator of the system of interest, $\rho(t)$. Thus, we first compute the reduced density operator $\rho(t)$ that corresponds to $W(q, p, t)$. The reduced density operator is presented below in a factorized operator form that makes evident the underlying physics and that is conducive to a study of positivity.

To obtain $\rho(t)$ from the Equation (5) for $W(q, p, t)$, the following steps are useful:

a) change the variables of integration according to

$$\begin{pmatrix} r \\ r' \end{pmatrix} = S^\dagger \begin{pmatrix} k \\ k' \end{pmatrix}, \tag{8}$$

where $S$ is the orthogonal matrix that satisfies

$$\begin{pmatrix} \lambda_+ & 0 \\ 0 & \lambda_- \end{pmatrix} = S^\dagger \begin{pmatrix} \omega X/\hbar & \dot{X}/2\hbar \\ \dot{X}/2\hbar & Y/\hbar\omega \end{pmatrix} S, \tag{9}$$

to diagonalize the last exponent in Equation (5):

$$\frac{\omega X}{\hbar}k^2 + \frac{Y}{\hbar\omega}k'^2 + \frac{\dot{X}}{\hbar}kk' = \lambda_+ r^2 + \lambda_- r'^2, \tag{10}$$

the eigenvalues $\lambda_+$ and $\lambda_-$ being given by [11]

$$\lambda_\pm = \frac{1}{2}\left[c + a \pm \sqrt{(c-a)^2 + b^2}\right] \geq 0 \tag{11}$$

where $a = \omega X/\hbar$, $b = \dot{X}/\hbar$ and $c = Y/\hbar\omega$, and the orthogonal matrix $S$ being chosen as

$$S = \frac{1}{\sqrt{2}} \begin{pmatrix} \dfrac{b}{\left[(c-a)^2 + b^2 + (c-a)\sqrt{(c-a)^2+b^2}\right]^{1/2}} & \dfrac{b}{\left[(c-a)^2 + b^2 - (c-a)\sqrt{(c-a)^2+b^2}\right]^{1/2}} \\ \dfrac{c-a+\sqrt{(c-a)^2+b^2}}{\left[(c-a)^2 + b^2 + (c-a)\sqrt{(c-a)^2+b^2}\right]^{1/2}} & \dfrac{c-a-\sqrt{(c-a)^2+b^2}}{\left[(c-a)^2 + b^2 - (c-a)\sqrt{(c-a)^2+b^2}\right]^{1/2}} \end{pmatrix}$$

(12)

if $b \geq 0$, and the same matrix but with the first column multiplied by $-1$ if $b < 0$;

b) letting $S_{ij}$ be the matrix elements of $S$, introduce the unitary operator $\tilde{M}$ defined by $\tilde{M} = M/R^{1/2}$ [12], where $R = \sqrt{\dot{A}^2 - A\ddot{A}}$ and where $M$ gives rise to the following time-dependent linear transformation of the position and momentum operators

$$M^\dagger \begin{pmatrix} q \\ p \end{pmatrix} M = \begin{pmatrix} \dot{A}S_{12} + \ddot{A}S_{22}/\omega & AS_{12}/m + \dot{A}S_{22}/m\omega \\ m\omega \dot{A}S_{11} + m\ddot{A}S_{21} & \omega AS_{11} + \dot{A}S_{21} \end{pmatrix} \begin{pmatrix} q \\ p \end{pmatrix}; \quad (13)$$

c) to introduce operators $\exp\left(-\dfrac{m\omega\lambda_+}{2\hbar}\{q,\cdot,q\}\right)$ and $\exp\left(-\dfrac{\lambda_-}{2\hbar m\omega}\{p,\cdot,p\}\right)$, where generally $\{A,\rho,B\} \equiv BA^\dagger \rho + \rho BA^\dagger - 2A^\dagger \rho B$, use the following identity [13]:

$$\exp(-\xi\{B,\cdot,B\})\rho = \frac{1}{2\pi}\int_{-\infty}^{\infty} du \sqrt{\frac{\pi}{\xi}}\, e^{-u^2/4\xi} V_B(u)\rho \quad (14)$$

where $\xi > 0$, $B$ is a Hermitian operator, and $V_B(u)\rho = (e^{-iuB}\rho e^{iuB} + e^{iuB}\rho e^{-iuB})/2$;

d) introduce a unitary operator $N$ satisfying

$$\begin{pmatrix} Q \\ P \end{pmatrix} = N \begin{pmatrix} q \\ p \end{pmatrix} N^\dagger = \begin{pmatrix} S_{12} & S_{22}/m\omega \\ m\omega S_{11} & S_{21} \end{pmatrix} \begin{pmatrix} q \\ p \end{pmatrix}; \quad (15)$$

and finally,

e) use the operator relation

$$\exp[i\tau(\{q,\cdot,p\}-\{p,\cdot,q\})]\rho(q,p) = \exp(4\hbar\tau)\rho[q\exp(2\hbar\tau), p\exp(2\hbar\tau)], \quad (16)$$

where $\tau$ is an arbitrary real parameter having the same dimensions as $\hbar^{-1}$.

The central result of this section follows:
$$\rho(t) = \exp\left[\frac{\ln R}{2i\hbar}(\{q,\cdot,p\}-\{p,\cdot,q\})\right] \times$$
$$\exp\left(-\frac{m\omega\lambda_+}{2\hbar R^2}\{Q,\cdot,Q\}\right)\exp\left(-\frac{\lambda_-}{2m\hbar\omega R^2}\{P,\cdot,P\}\right)(N\tilde{M}\rho(0)\tilde{M}^\dagger N^\dagger) \quad (17)$$

In this exact operator form, the meaning of each of the factors is manifest. In particular, the first exponential factor gives rise to dissipation, the second and third factors give rise to fluctuations arising from the interaction of the reservoir with the system of interest, and the final factor involving the unitary operators $N$ and $\tilde{M}$ gives rise to reversible evolution.

**Outer solution**

The generator $L$ of Equation (1) may be obtained from Equation (5) by i) differentiating with respect to time, ii) invoking a high temperature approximation, and iii) considering the *outer limit* where $\alpha \to \infty$ with $t$ fixed but not equal to zero [2]. The propagator $\exp(tL)$ approximates the outer solution of the dynamics of the harmonic oscillator.

The last limit is delicate: as a consequence of the fact that the $\alpha \to \infty$ limit is not uniform in time, a boundary layer arises that renders the outer solution invalid for times shorter than $O(1/\alpha)$. Because of this non-uniformity, the $t \to 0$ limit of the outer solution, referred to as a "slip" or "effective initial data" [2], does not correspond to the true initial condition [14]. For times shorter than $\Delta t = O(1/\alpha)$, an inner solution, describing the rapid entanglement of the system of interest and the reservoir, should instead be considered, and is the key to preserving positivity.

**Inner solution**

To obtain the inner solution associated with Equation (17), we consider the *inner limit* in which $\alpha \to \infty$ with $\alpha t$ constant but non-zero. Since in the inner limit, $R(t) \sim 1$, Equation (17) simplifies to

$$\rho(t) = \exp\left(-\frac{m\omega}{2\hbar}\lambda_+\{Q,\cdot,Q\}\right)\exp\left(-\frac{1}{2\hbar m\omega}\lambda_-\{P,\cdot,P\}\right)(NM\rho(0)M^\dagger N^\dagger) \quad (18),$$

In the inner limit, we also have $A \sim t$, $\dot{A} \sim 1$, $\ddot{A} \sim 2\Gamma(e^{-\tilde{\alpha}} - 1)$,

$$S \sim \begin{pmatrix} \omega t/2 & 1 \\ 1 & -\omega t/2 \end{pmatrix}, \quad (19)$$

$$\lambda_+(\tilde{\alpha},\alpha) \sim \frac{4\Gamma}{\beta_{max}\omega}\left(\exp(-\tilde{\alpha}) + \tilde{\alpha} - 1\right) \geq 0, \quad (20)$$

and

$$\lambda_-(\tilde{\alpha},\alpha) \sim \frac{2\Gamma\omega}{\beta_{max}}\left(\frac{\tilde{\alpha}}{\alpha}\right)^2\left[\frac{\tilde{\alpha}}{6} - \frac{1}{2} - \frac{\exp(-\tilde{\alpha})}{2} - \frac{2\exp(-\tilde{\alpha})}{\tilde{\alpha}} + \frac{2[1-\exp(-\tilde{\alpha})]}{\tilde{\alpha}^2}\right] \geq 0 \quad (21)$$

where $\tilde{\alpha} = \alpha t$ with $t > 0$, and $\beta_{max} = \hbar\alpha/kT$ [15].

The inner solution (18) is manifestly positive, involving as it does either unitary operators, or operators of the form $\exp(-\{B,\cdot,B\})$ [16]. What is not obvious, however, but what we now show, is that propagation of *all* density operators with the outer propagator yields a positive operator provided the inner propagator acts first: $\exp[(t - \Delta t)L] \, J(\Delta t)\rho(0) \geq 0$ where $0 \leq \Delta t = O(1/\alpha)$ [10].

There is no inconsistency here. The propagator $\exp(tL)$, not being of Lindblad form [16] (i.e., the exponent is not a sum of terms of the form $-\{B,\cdot,B\}$), does not preserve positivity on the set V. However, as we show below, on the smaller set V' = J(Δt)V, it does preserve positivity.

**Positivity**

We now show that under appropriate conditions stated below
$$\exp[(t-\Delta t)L] \, J(\Delta t)\rho(0) \geq 0. \quad (22)$$
For this purpose, we introduce the following operators:
$L_H = a[q^2,\cdot] + b[p^2,\cdot] - \gamma[pq+qp,\cdot]$; $G_1 = a\{q,\cdot,q\} + b\{p,\cdot,p\}$; $G_2 = a\{q,\cdot,q\} - b\{p,\cdot,p\}$;
$G_3 = \{p,\cdot,q\} + \{q,\cdot,p\}$; and $G_4 = \{q,\cdot,p\} - \{p,\cdot,q\}$, where $a = \frac{m\omega^2}{2\hbar i}$, $b = \frac{1}{2m\hbar i}$ and $\gamma = \frac{\Gamma i}{2\hbar}$.

We have $\exp(tL) = \exp\left[tL_H + t\gamma\left(G_4 - \frac{4kT}{\omega^2}G_1 - \frac{4kT}{\omega^2}G_2\right)\right]$. These operators are useful because the set $\{L_H, G_1, G_2, G_3, G_4\}$ forms a closed algebra under commutation, with multiplication table

|       | $L_H$ | $G_1$ | $G_2$ | $G_3$ | $G_4$ |
|-------|-------|-------|-------|-------|-------|
| $L_H$ | 0 | $4\hbar i\gamma G_2$ | $4\hbar i(\gamma G_1 - abG_3)$ | $4\hbar iG_2$ | 0 |
| $G_1$ |   | 0 | 0 | 0 | $-4\hbar iG_1$ |
| $G_2$ |   |   | 0 | 0 | $-4\hbar iG_2$ |
| $G_3$ |   |   |   | 0 | $-4\hbar iG_3$ |
| $G_4$ |   |   |   |   | 0 |

(for example, $[L_H, G_1] = 4\hbar i\gamma G_2$). Because the algebra is closed, we may use the method of Wei and Norman [17] to write the following equation that enters into expression (22):

$$\exp(t_2 L)\exp\left(-\frac{m\omega\lambda_-(t_1)}{2\hbar}\{Q(t_1),\cdot,Q(t_1)\}\right)\exp\left(-\frac{\lambda_+(t_1)}{2\hbar m\omega}\{P(t_1),\cdot,P(t_1)\}\right) = \exp(t_2 L_H) \times$$

$$\exp\{\gamma t_2[4\hbar ia(E(t_2)+C(t_2))\{q,\cdot,q\} + 4\hbar ib(E(t_2)-C(t_2)-i\lambda(t_1)/\omega)\{p,\cdot,p\} +$$

$$(e^{-4\hbar i\gamma t_2} - 1 + 4\hbar iD(t_2))\{p,\cdot,q\} + (e^{-4\hbar i\gamma t_2} - 1 + 4\hbar iD(t_2))\{q,\cdot,p\}]/(1-e^{-4\hbar i\gamma t_2})\} \times$$

$$\exp-\left\{\sqrt{|2as_2(t_1)|}q - \frac{s_1(t_1)}{\sqrt{|2as_2(t_1)|}}p, \cdot, \sqrt{|2as_2(t_1)|}q - \frac{s_1(t_1)}{\sqrt{|2as_2(t_1)|}}p\right\} \quad (23)$$

where

$$\lambda(t) = \frac{\lambda_+(t)\lambda_-(t)}{S_{11}^2(t)\lambda_-(t) + S_{12}^2(t)\lambda_+(t)}, \quad (24)$$

$$s_1(t) = -(S_{11}(t)S_{21}(t)\lambda_-(t) + S_{12}(t)S_{22}(t)\lambda_+(t))/2\hbar, \quad (25)$$

$$s_2(t) = -\frac{m\omega}{4\hbar a}(S_{11}^2(t)\lambda_-(t) + S_{12}^2(t)\lambda_+(t)), \text{ and}$$

$$(C(t),D(t),E(t)) = \alpha e^{-4\hbar i\gamma t}\left(\sqrt{\gamma^2-ab}\sin(4\hbar t\sqrt{\gamma^2-ab}), iab(1-\cos(4\hbar t\sqrt{\gamma^2-ab})),\right.$$

$$(i\gamma)^{-1}\left[ab(1-e^{4\hbar i\gamma t}) + \gamma^2(e^{4\hbar i\gamma t} - \cos(4\hbar t\sqrt{\gamma^2-ab}))\right]\bigg)/[4\hbar(\gamma^2-ab)]. \quad (26)$$

In view of expression (22), we will ultimately take $t_2$ to be equal to $t - \Delta t$ and $t_1$ to be $\Delta t = O(1/\alpha)$. On the right hand side of Equation (23), the first exponential factor $\exp(t_2 L_H)$ is associated with unitary evolution and clearly preserves positivity. On account of identity (14), the third exponential factor also preserves positivity. The second exponential factor of Equation (23) can be shown [18] to preserve positivity if $\omega > \Gamma$ and if

$$\left| \frac{e^{-4\hbar i \eta_2} - 1}{4\hbar i} + D(t_2) \right|^2 \leq ab \big( E(t_2) + C(t_2) \big) \big( E(t_2) - C(t_2) - 2i\lambda(t_1)/\omega \big), \tag{27}$$

which can indeed be shown to be true in a high temperature regime defined by:

$$\frac{9}{2} \left( \frac{kT}{\hbar\omega} \right)^2 \left[ \left( \frac{2kT}{\hbar\omega} \right)^2 - 1 \right] \lambda^2(t_1) > 1, \tag{28}$$

where we note that in the inner limit $\lambda(t_1) \sim \lambda_-(t_1)$, the last quantity being given by Equation (21) whose right hand side is an increasing function of time. We conclude that when $\omega > \Gamma$ and inequality (28) is satisfied, expression (22) is true.

As a global propagator valid for all times $t \geq \Delta t = O(1/\alpha)$, the left hand side of expression (22) is rather naïve in that we have simply patched the inner and outer limits. (For times $t < \Delta t$, $\rho(t) = J(\Delta t)\rho(0)$.) A better approach, which we are currently investigating, employs boundary layer theory. Either way, the physics is clear. Being initially uncorrelated, the harmonic oscillator and the reservoir become rapidly entangled on a time scale $O(1/\alpha)$. This rapid dynamic behavior, which is associated with a boundary layer, is governed by the positivity-preserving inner propagator of Equation (18). This propagator maps the set of all density operators V into the smaller set V′, on which exp($tL$) preserves positivity under the conditions stated above.

V′ is a proper subset of V because pure density operators are generally mapped into mixed states, with an accompanying rise in entropy. For example, it can be shown that the linear entropy $-k \ln \text{Tr} \rho^2(t)$, with $\rho(t) = J(t)[M(t)^\dagger \rho(0) M(t)] \in$ V′ and with $\rho(0)$ being the coherent state $|0\rangle\langle 0|$, is given by

$$\frac{k}{2} \ln \big[ (2\lambda_+(t) + 1)(2\lambda_-(t) + 1) \big]. \tag{29}$$

From the inner limit expressions (20) and (21), one can show that $\lambda_+(t), \lambda_-(t)$ and the linear entropy are increasing functions of time (although Equation (29), like all inner limit expressions, are only valid for $t = O(1/\alpha)$).

It is also of interest to note that for *any* element of V′, the position and momentum uncertainties must satisfy

$$\Delta q^2 \geq \frac{\hbar}{m\omega}(\lambda_- S_{21}^2 + \lambda_+ S_{22}^2) \sim \frac{2\Gamma kT}{m\alpha} \left( \frac{\tilde{\alpha}}{\alpha} \right)^2 \left[ \frac{2\tilde{\alpha}}{3} - 1 - \frac{2\exp(-\tilde{\alpha})}{\tilde{\alpha}} + \frac{2[1 - \exp(-\tilde{\alpha})]}{\tilde{\alpha}^2} \right] \tag{30}$$

$$\Delta p^2 \geq m\hbar\omega(\lambda_- S_{11}^2 + \lambda_+ S_{12}^2) \sim \frac{4\Gamma mkT}{\alpha}\left(\tilde{\alpha} + \exp(-\tilde{\alpha}) - 1\right), \tag{31}$$

the asymptotic expressions being valid in the inner limit.

The present study sheds light on a previous approach [4] for obtaining positive evolution that considered an initially correlated total density operator, instead of the uncorrelated state assumed in the above analysis. $J(\Delta t)\rho(0)$ corresponds to $\text{Tr}_r \rho_T(\Delta t)$, the trace over reservoir variables of the total density operator evaluated at $\Delta t$. If we were to take $\Delta t$ as our initial time, with $\rho_T(\Delta t)$ the total initial correlated state, then we have seen from the above analysis that the appropriate inner limit of $\text{Tr}_r \rho_T(\Delta t)$ remains positive when propagated with the conventional quantum Brownian generator $L$. Starting with an initially correlated state is a promising approach because, as shown above, it negates the necessity of having to employ a propagator of Lindblad form to preserve positivity [19].

Finally, we note that in a numerical study of the Markovian Redfield equations describing a two-level system coupled to a reservoir, Suarez et al. [9] observe that non-local effects seem to restrict a class of density matrices that can evolve under the Redfield equations, and that only this restricted class has to be mapped into a set of positive matrices. Noting that the slipped or effective initial data is related to the fast adjustment of the system and reservoir to each other, Suarez et al. suggest that positivity would be preserved under the Redfield equations if the effective initial data were used. However, for the harmonic oscillator investigated here, the use of the aforementioned effective initial data is not a general remedy to cure the positivity problem. For example, in Ref. [2], it is shown that the effective initial data are not generally positive. In contrast, we have shown that the set $J(\Delta t)\ V$ is positive and, when $\omega > \Gamma$ and expression (28) hold, that the conventional quantum Brownian propagator preserves positivity thereon.

-------------

[11] The eigenvalues $\lambda_\pm$ can be shown to be non-negative by using Equations (6) and (7), and the Schwartz inequality.

[12] We assume $R > 0$, which is true when $2\Gamma(1 + 2\Omega/\lambda + \Gamma^2/\lambda^2) < \Omega$. This last condition, together with the aforementioned dependence of $S$ on the sign of $b$ (which ensures that the determinant of the 2x2 matrix of Equation (13) is +1), guarantees that $M$ is a metaplectic operator. Metaplectic operators are unitary operators giving rise to linear transformations of $(q, p)$. See, for example, R. G. Littlejohn, Phys. Rep. **138**, 193 (1986).

[13] This identity may be proved by considering a Hermitian operator $B$, by defining $V_B(u) = (e^{-iuB} \cdot e^{iuB} + e^{iuB} \cdot e^{-iuB})/2$, with Fourier transform $\hat{V}_B(k) = \frac{1}{\sqrt{2\pi}} \int_{-\infty}^{\infty} du\, e^{iuk} V_B(u)$, and by noting that $\{B, \cdot, B\} = [B, [B, \cdot]]$, which allows us to write, upon expanding $V_B(u)$ as an infinite series, the following:

$$\exp(-\xi\{B, \cdot, B\}) = \sum_{n=0}^{\infty} \frac{\xi^n}{n!} \int_{-\infty}^{\infty} du\, \delta(u) \frac{d^{2n}}{du^{2n}} V_B(u)$$

$$= \frac{1}{\sqrt{2\pi}} \int_{-\infty}^{\infty} dk\, \hat{V}_B(k) \sum_{n=0}^{\infty} \frac{(-\xi k^2)^n}{n!}$$

$$= \frac{1}{2\pi} \int_{-\infty}^{\infty} du\, \sqrt{\frac{\pi}{\xi}} e^{-u^2/4\xi} V_B(u),$$

where $\xi > 0$. For $B$ Hermitian, this identity shows that $\exp(-\xi\{B, \cdot, B\})\rho$ is positive. (In a similar manner, the operator $\exp(\xi\{B, \cdot, B\})$ may be shown to be equal to the right hand side of Equation (14), but with $V_B(u)$ replaced with $V_B(iu)$; the operator $\exp(\xi\{B, \cdot, B\})$ does not generally preserve positivity on the set of density operators, V.) A variant of Equation (14) was also derived in a different manner in N. G. van Kampen, J. Stat. Phys. **78**, 299 (1995); note, however, that some factors are incorrect in the expression for $B_\alpha$ in the Appendix of this last reference.

[14] $\exp(tL)$ alone is an approximation of the outer limit propagator because the slip is neglected.

[15] To obtain these eigenvalues, we first considered a high temperature expansion in the parameter $\beta_{max}$, followed by considering the limit $\alpha \to \infty$ with $\tilde{\alpha}$ fixed.

[16] Propagators of the form $\exp(-\sum_\alpha \{B_\alpha, \cdot, B_\alpha\})$, where the $B_\alpha$ are not necessarily Hermitian, were shown to preserve positivity in G. Lindblad, Commun. Math. Phys. **48**, 119 (1976). That $\exp(-\{B, \cdot, B\})$ preserves positivity, with $B$ Hermitian, may also be gleaned from Equation (14).

[17] J. Wei and E. Norman, Proc. Am. Math. Soc. **15**, 327 (1964).

[18] Suppose $L = -(A\{q, \cdot, q\} + B\{p, \cdot, p\} + C\{q, \cdot, p\} + D\{p, \cdot, q\})$, where $A, B, C, D$ are complex numbers. It follows that complex vectors $\vec{a} = (a_1, a_2, ...)$ and $\vec{b} = (b_1, b_2, ...)$ exist

such that $L = -\sum_j \{a_j q + b_j p, \cdot, a_j q + b_j p\}$ if, and only if, $A = \vec{a} \cdot \vec{a}$, $B = \vec{b} \cdot \vec{b}$, $C = \vec{a} \cdot \vec{b}$, and $D = \vec{b} \cdot \vec{a}$. From these expressions, and the Schwartz inequality, it follows that $L$ may be written in this latter form if, and only if, $A \geq 0, B \geq 0, C = D^*$ and $AB \geq |C|^2$. This result was proved by J.E. Sipe (unpublished). Expression (27) corresponds to this last inequality.

[19] In this vein, see also P. Pechukas, Phys. Rev. Lett. **73**, 1060 (1994).